\documentclass[
 amsmath,amssymb,
 pra,twocolumn
]{revtex4-2}

\usepackage{graphicx}
\usepackage{dcolumn}
\usepackage{bm}
\usepackage[colorlinks=true, citecolor=blue, linkcolor=blue, urlcolor=blue]{hyperref}
\usepackage{appendix}
\usepackage{braket}
\usepackage{ stmaryrd }
\usepackage{float}
\usepackage{xspace}
\usepackage{acronym} 
\usepackage{physics}
\usepackage{soul}
\graphicspath{ {Figures/} }

\newcommand{\kett}[2][+]{\ket{#1\frac{#2}{2}}}

\newcommand{\dmeq}[1]{$\Delta m_I = {#1}$}
\newcommand{\tplus}{$^{3+}$\xspace}
\newcommand{\ErYSO}{$^{167}$Er$^{3+}$:Y$_2$SiO$_5$\xspace}

\newcommand{\Er}{$^{167}$Er}
\newcommand{\dm}{\ensuremath{\Delta m_I}\xspace}
\newacro{ecdl}  [ECDL]  {External Cavity Diode Laser}
\newacro{eom}  [EOM]  {Electro-Optic Modulator}
\newacro{afc}  [AFC]  {Atomic Frequency Comb}
\newacro{gem}  [GEM]  {Gradient Echo Memory}
\newacro{od}  [OD]  {Optical Depth}
\newacro{rose}  [ROSE]  {Revival of Silenced Echo}

\begin{document}

\preprint{APS/123-QED}

\title{Initialisation protocol for efficient quantum memories using resolved hyperfine structure}

\author{James Stuart}
\author{Morgan Hedges} 
\author{Rose Ahlefeldt}
\author{Matthew Sellars}
\affiliation{Centre for Quantum Computation and Communications Technology, Research School of Physics, Australian National University, Canberra, ACT, Australia}
\date{\today}

\begin{abstract}

We describe a quantum memory spectral preparation strategy that optimises memory efficiency and bandwidth in materials such as \ErYSO in a high field regime, where the hyperfine structure is resolved. We demonstrate the method in \ErYSO by preparing spectrally isolated 18~dB-absorbing features on a $<1$~dB background. Using these features we create an atomic frequency comb and show  quantum storage of 200 ns pulses  with 22\% efficiency, limited by the background absorption which arises from laser instability. We describe the experimental improvements needed to reach the material limits: $\order{1}$~s spin state storage, $\order{100}$~MHz bandwidth, and $>90$\%  efficiency.
\end{abstract}

\maketitle

Quantum memories are an important component in most applications of quantum information, including both large-scale quantum computing and long-distance secure communication\cite{tittel2010,Felix2013}. Such memories can be created using ensembles of rare earth ions doped into crystals. These systems have shown high potential storage density\cite{storageDensity,QMdataCapacity,Jobez2016}, high potential efficiency when enhanced with an optical cavity \cite{Simon2010}, and very long potential storage times\cite{Grace2014, Ma2021, Holz2020}.

For quantum communication applications, Er\tplus:Y$_2$SiO$_5$ has attracted particular attention. The $^4$I$_{15/2} \shortrightarrow$ $^4$I$_{13/2}$ transition of Er\tplus lies within the low loss telecom band at 1550 nm. In Y$_2$SiO$_5$, a magnetic field of $\order{7\mbox{ T}}$ and a temperature below 2~K freezes out the Er\tplus spin flips that dominate dephasing, resulting in long lifetimes and coherence times on the optical transition ($\sim10$~ms lifetime and $\sim4$~ms coherence time) \cite{BottgerLifetime} as well as on the $I = 7/2$ hyperfine transitions of \Er\tplus, the only isotope with nuclear spin (10~min lifetime and 1.3~s coherence time)\cite{ Rancic2016}. These values are comparable to Pr$^{3+}$ and Eu$^{3+}$, the non-Kramers ions typically used to demonstrate high performance quantum memories to date\cite{Grace2014,hedges10,schraft16}, suggesting comparable or improved memory performance is achievable in the telecom band using $^{167}$Er$^{3+}$. 

The key requirements for global-scale quantum repeater applications of quantum memories are $\order{100}$~ms storage times and efficiencies above 90\% \cite{tittel2010, Razavi2009}.  A high data rate is also desired, implying a large memory bandwidth. These requirements place restrictions on materials used for memories. Long storage times can be achieved using a spin-state storage quantum memory protocol as long as hyperfine coherence time is long. Efficiency depends on the spin storage protocol used, but the base requirement is a high absorption on the ions participating in the memory and low absorption on any spectator (non-participating) ions. Bandwidth is determined by the maximum width of a transmissive window that can be created at the storage frequency, since spin-state storage protocols all require resonant spectator ions (those not participating in the memory) to be shifted to a non-resonant state.

Therefore, the memory performance in \ErYSO is optimised by creating high-contrast spectral features well resolved from any spectator ions. Here, we show how to prepare such features. We use the method to prepare an atomic frequency comb (AFC) in the system and show non-classical delay of coherent states. We did not demonstrate the final step (spin state storage) as an appropriate light source for the shelving pulses was not available; instead, we describe this step and other experimental modifications required to reach the full system performance. We consider \ErYSO here, but the method is applicable to other crystals with similar resolved hyperfine structures and lifetimes, such as other Kramers ions at high fields and low temperatures.

\begin{figure}
        \centering
        \includegraphics[width = 0.95\linewidth]{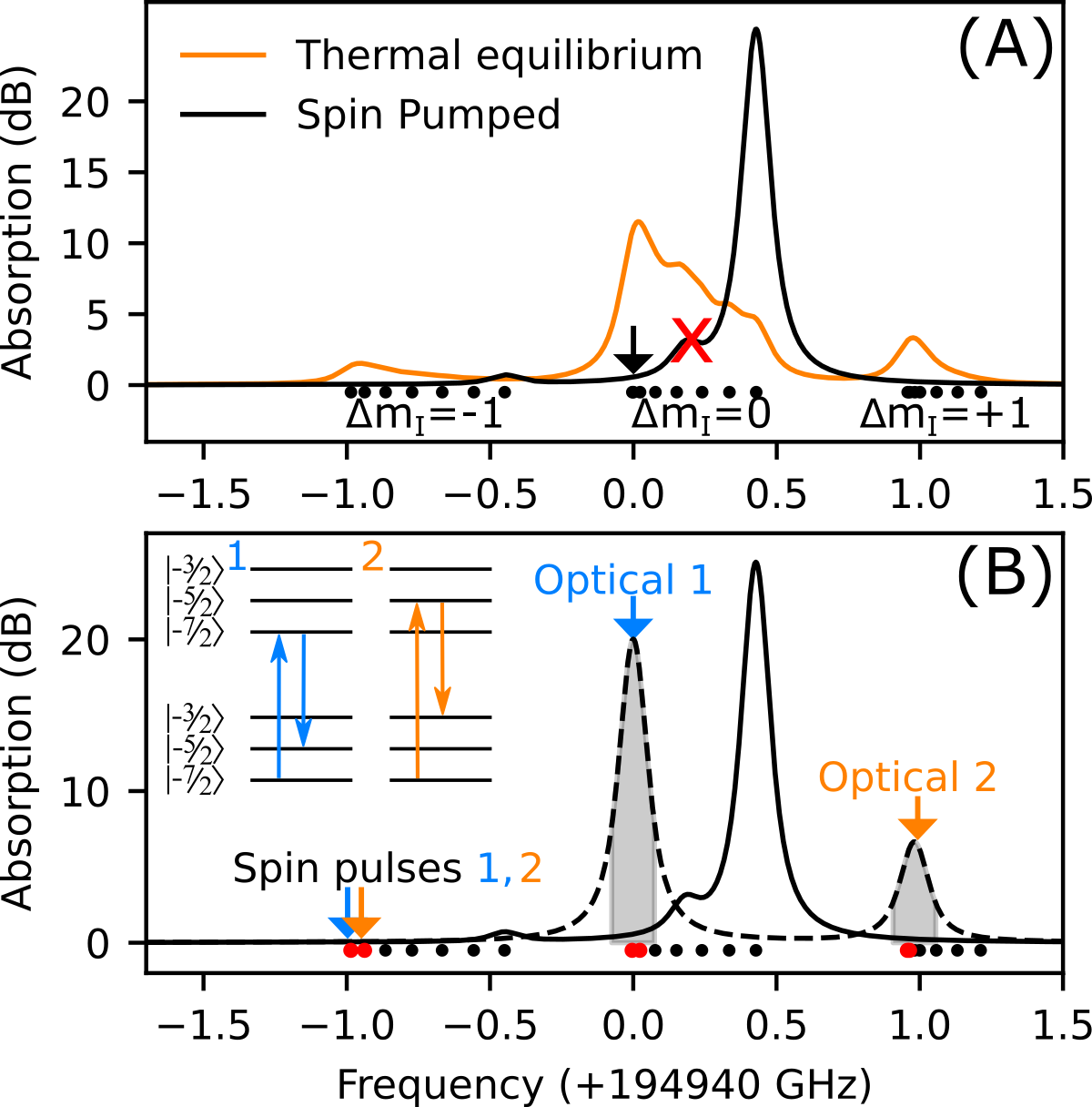}
        \caption{(\textbf{A}). Absorption spectrum of 0.005\% \ErYSO (92\% $^{167}$Er$^{3+}$ isotopic purity) before (orange) and after (black) spin polarisation, with peaks labelled according to $\Delta m_I$. Arrow: location of AFC created, cross: location of the 8\% $I=0$ impurity line.  (\textbf{B}) Solid line: spin polarised spectrum from (A) . Dashed line: simulation of the  absorption spectrum of $\kett[-]{7}_g$, with potential memory bandwidth and location (shaded gray areas) and possible spin storage $\Lambda$ systems indicated (see text). Arrows and red dots indicate driven transitions for each $\Lambda$ system. }
        \label{fig:spin_pumped}
    \end{figure}

We consider \ErYSO in a large field, where the electron spin is frozen and the hyperfine lifetime and coherence time are long. In particular, we use the experimental regime of Ref. \cite{Rancic2016}: 1.5~K and a field of 7~T applied  along the $D_1$ optical extinction axis. Fig. \ref{fig:spin_pumped} (A) shows the thermal equilibrium absorption spectrum (orange line) of the $^4$I$_{15/2} \shortrightarrow$ $^4$I$_{13/2}$ transition of site 2\cite{BottgerLifetime}. The spectrum consists of three bands of peaks corresponding to optical transitions with nuclear spin projection $\Delta m_I = m_I(g)-m_I(e) = -1, 0, +1$, with transitions in each peak mostly ordered according to $m_I(g)$. 

The first step in the spectral preparation process is to spin polarise the ensemble into one of these extreme hyperfine states ($m_I(g)=+\frac{7}{2}$ ($m_I(g)=-\frac{7}{2}$)), by sweeping a laser over the $\Delta m_I = +1$ ($\Delta m_I = -1$) band (black line in Fig \ref{fig:spin_pumped} (A)). The cross in this figure marks the extra absorption due to $I=0$ Er$^{3+}$ isotope impurities, present at 8\% in the sample studied, which cannot be affected by spectral preparation. The next step is to transfer a spectral sub-ensemble (memory ions) back to the other extreme hyperfine state.  Using the two extreme states ensures maximal spectral separation of memory and bulk ions (all ions not participating in the memory), maximising the transmissive window the memory ions sit in and therefore the memory bandwidth. The large detuning between bulk and memory ion spins also suppresses cross-relaxation and increases the memory spin lifetime, compared to a non-polarised ensemble. Of the extreme hyperfine states, the $\ket{-\frac{7}{2}}_g$ level is better suited for the memory ions since the oscillator strengths of the $\abs{\Delta m_I} > 0$ transitions required for spin-state storage increase with decreasing $m_I(g)$.  

In Fig. \ref{fig:spin_pumped} (B) we have simulated the spectrum of the $\ket{-\frac{7}{2}}_g$ level to illustrate the possible locations and bandwidth of the memory prepared using this level. We also show possible $\Lambda$ systems suitable for spin state storage. To maximise the oscillator strengths and, thus, potential memory efficiency, the choices of $\Lambda$ systems are restricted to those using $\Delta m_I = 0,\pm 1$ transitions. Further, control pulses must be applied on $\Delta m_I=-1$, because the memory bandwidth is larger than the separation of adjacent optical transitions, so control pulses on $\Delta m_I = 0,+1$ overlap with, and so will excite, transitions from the $\ket{-\frac{7}{2}}_g$ state. There are, then, two possible $\Lambda$ systems: 1)$\ket{-\frac{7}{2}}_g\rightarrow\ket{-\frac{7}{2}}_e$ ($\Delta m_I=0$) for optical storage and $\ket{-\frac{5}{2}}_g\rightarrow\ket{-\frac{7}{2}}_e$ for spin state storage; and 2)$\ket{-\frac{7}{2}}_g\rightarrow\ket{-\frac{5}{2}}_e$ ($\Delta m_I=+1$) for optical storage and $\ket{-\frac{3}{2}}_g\rightarrow\ket{-\frac{5}{2}}_e$ for spin state storage. These options are illustrated in Fig. \ref{fig:spin_pumped} (B). Option 1) has a higher oscillator strength on the optical storage transition, more suitable for a free-space implementation such as that shown here. Option 2) has only 30\% of the peak oscillator strength of option 1) but has a much lower background absorption (6\% of option 1)) since the $\Delta m_I=+1$ transition is spectrally much further away from the spin-polarised bulk as well as the absorption from Er$^{3+}$ isotopic impurity ( all $I=0$). It may be suitable for the cavity implementations, discussed at the end this paper, where peak optical depth does not limit the memory efficiency. However, in the paper we demonstrate option 1).

We spin polarise the bulk of the ions into $ \ket{+ \frac{7}{2}}_g$ and selectively anti-polarise ions into $ \ket{- \frac{7}{2}}_g$. These memory ions, addressed via their $\Delta m_I=0$ transition, are used to demonstrate quantum storage with the AFC protocol. The simple extension to spin-state storage on the $ \ket{- \frac{5}{2}}_g$ ground state could be achieved using an optical $\pi$-pulse at a detuning of 997~MHz immediately after the pulse is applied. 

The experimental setup is similar to previous work \cite{Rancic2016}. The sample is a 0.005\% \ErYSO crystal (isotopic purity of 91.77\%), mounted in a helium bath cryostat containing a superconducting magnet to supply a field of 7~T vertically. Optical access to the sample is provided by a single window on the bottom of the cryostat. The crystal is mounted with the $D_1$ axis vertical and the $D_1$ face against a mirror, such that the laser beam double-passes the crystal.  Two laser sources  are combined in-fiber for the input beam: for spin polarisation, a broadly-scannable Pure Photonics PPLC300 laser; and for spectral preparation and pulse storage, a Thorlabs FPL1009S external-cavity diode laser locked to a home-built fiber reference cavity to a sub-millisecond linewidth of 20~kHz, but is sensitive to acoustic noise on longer timescales (discussed below). The stabilised laser is amplitude-modulated with an electro-optic modulator, with the upper sideband driving the $\dm=-1,0$ bands. The (suppressed) carrier is located $-1.5$~GHz from the center of the $\Delta m_I= 0$ transition, so both the carrier and low-frequency sideband experience minimal absorption. The output beam is spatially separated from the input beam below the cryostat and collected in a balanced heterodyne detection system, for which the local oscillator is derived from the stabilised laser.

The entire ensemble was spin polarised into the $\ket{+\frac{7}{2}}_g$ level by sweeping over the $\Delta m_I = +1$ band for 10~s at a rate of 25~Hz with 3~mW of power and a beam waist of $40\pm 1$~$\mu$m. This pumped 95\% of the ions into the $\ket{+\frac{7}{2}}_g$ level and raised the peak optical depth from 4~dB to 25~dB. Next, a spectrally narrow subgroup of ions was transferred to the $\ket{-\frac{7}{2}}_g$ level by a series of short (100~$\mu s$) burns: first on the $\ket{\frac{7}{2}}_g\rightarrow\ket{\frac{3}{2}}_e$ transition in the $\Delta m_I = -2$ band, and then successively on the five transitions in the $\Delta m_I = -1$ band from $\ket{\frac{3}{2}}_g\rightarrow\ket{\frac{1}{2}}_e$ to $\ket{-\frac{5}{2}}_g\rightarrow\ket{-\frac{7}{2}}_e$. This anti-polarisation sequence was repeated 250 times (150~ms total time). The first burn was on a  $\Delta m_I = -2$ transition rather than a $\Delta m_I = -1$ transition to avoid population in the $\kett{5}$ level. This is because in the \dmeq{-1} line the $\ket{+\frac{5}{2}}_g$ overlaps with the tail of the spin-pumped inhomogeneous line. As a result, burning on the $\Delta m_I = -1$ $\kett[-]{5}$ line would have the undesirable effect of producing additional spectral structure.
   
    \begin{figure}
        \centering
        \includegraphics[width = 0.95\linewidth]{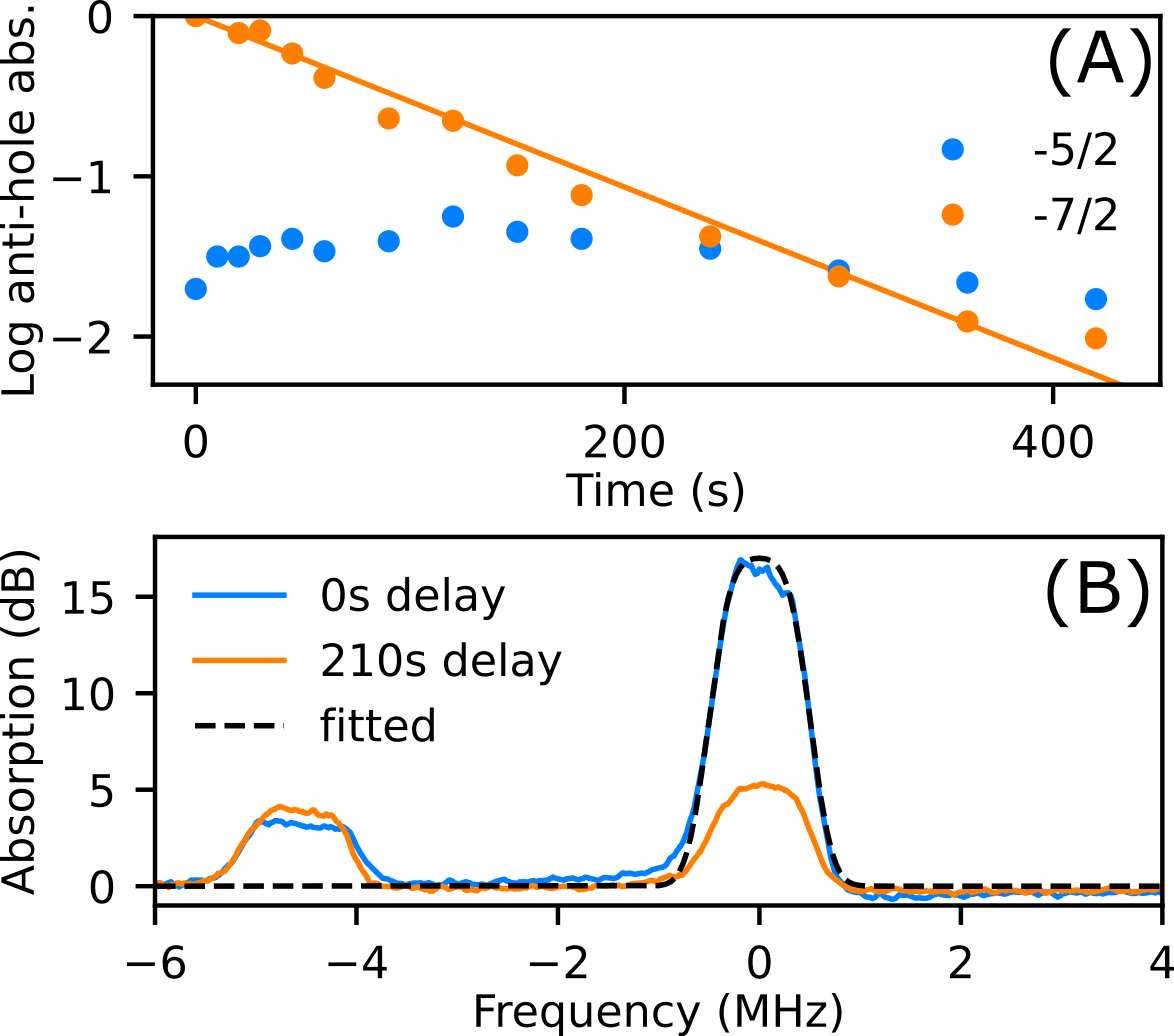}
        \caption{(\textbf{A}). Lifetime of the feature prepared in the $\kett[-]{7}_g$ hyperfine level with fit (orange) and the $\kett[-]{5}_g$ anti-hole. (\textbf{B}). Spectrum of the feature. The population in $\ket{-\frac{7}{2}}_g$ causes the feature at 0~MHz and residual $\ket{-\frac{5}{2}}_g$ population the feature at $-4.5$~MHz.  The dashed line shows a fit to the 0~MHz spectral feature.}
        \label{fig:antiholelifetime}
    \end{figure}
    
The system was probed with a strong pulse before and after the creation of the spectral feature to determine the absorption of the feature itself above any background absorption or optical losses in the path. The background at the frequency of the $\ket{-\frac{7}{2}}_g$ was separately measured to be $0.48\pm0.06$~dB by saturating a narrow band of ions before the feature was created and probing the resultant spectral hole. Using a model of the spin-polarised optical spectrum \cite{MilosThesis}, we estimate that 0.08~dB of this background arises from the tail of the $I=0$ impurity line at this optical frequency, 0.3~dB from the tail of the spin-polarised bulk line, and the remaining 0.1~dB from imperfect spin polarisation.

The resulting background-subtracted absorption of a 1 MHz spectral feature prepared in the $\ket{-\frac{7}{2}}_g$ level is shown in Fig. \ref{fig:antiholelifetime} (B).   Residual population in the  $\ket{-\frac{5}{2}}_g$ level is also visible at -4.5~MHz. Earlier we said that the optical frequency generally increases with $m_I(g)$; the $\ket{-\frac{5}{2}}_g$ and $\ket{-\frac{7}{2}}_g$ transitions in the $\Delta m_I=0$ band are one of the few exceptions to this trend. These features can be described by the convolution (dashed line) of the square excitation with a 0.4~MHz Gaussian broadening function. This broadening is dominated by low-frequency laser instability, although for short burn times can be limited by the 130~kHz inhomogeneous broadening on the hyperfine transitions\cite{MilosThesis}. 

The $\ket{-\frac{7}{2}}_g$ feature relaxes on the timescale of hundreds of seconds, as shown by the orange line in Fig. \ref{fig:antiholelifetime} (B). Some of the ions decay via a single-phonon process as was suggested by Ran\v{c}i\'{c} et al. \cite{Rancic2016}, with the population in the $\ket{-\frac{5}{2}}_g$ level initially increasing as it is pumped by the $\ket{-\frac{7}{2}}_g$. The lifetime of  the $\ket{-\frac{7}{2}}_g$ spectral feature is $188 \pm 8$~s (orange line in Fig. \ref{fig:antiholelifetime} (A)). As expected, this exceeds the 60~s lifetime previously measured for spectral holes burnt in a thermal equilibrium population\cite{Rancic2016}, where cross-relaxation dominated the lifetime. However, the lifetime is shorter than the spin-lattice relaxation time of 600~s calculated in \cite{Rancic2016}, suggesting that hyperfine cross-relaxation is not completely suppressed. These relaxation processes will be the subject of a subsequent paper.

    \begin{figure}
        \centering
        \includegraphics[width = 0.9\linewidth]{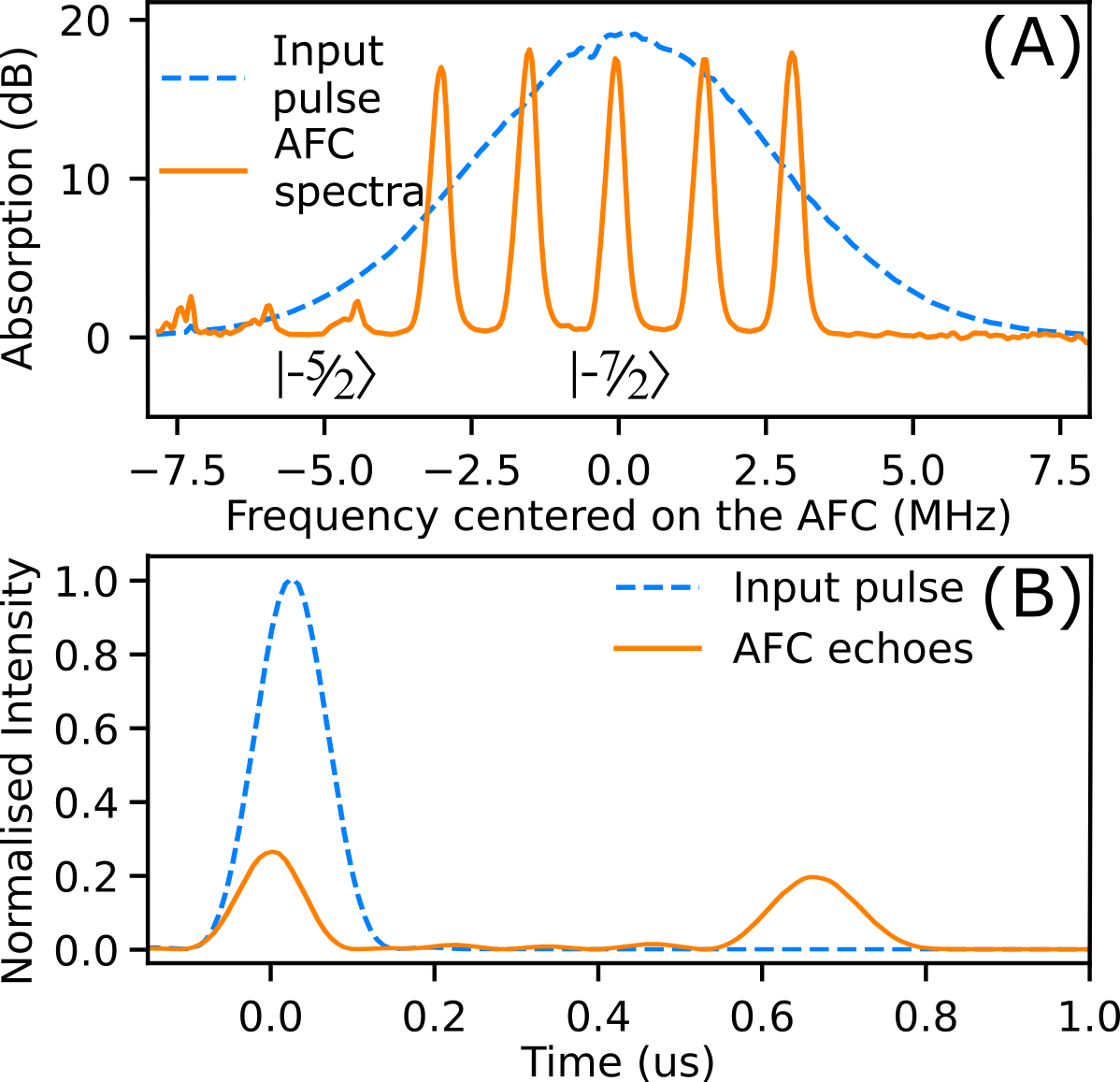}
        \caption{(\textbf{A}). Spectrum of the \ac{afc}  created in the $\ket{-\frac{7}{2}}_g$ level. Residual population in $\ket{-\frac{5}{2}}_g$ is visible below $-4$~MHz. Blue dashed trace: spectrum of the input pulse. (\textbf{B}). Time trace of an  \ac{afc} echo. Blue dashed line: input pulse sent through the ensemble before \ac{afc} creation. Orange line: the same input pulse sent through the \ac{afc}.}
        \label{fig:5toothAFC}
    \end{figure}

To demonstrate quantum storage, an AFC was created by applying the above preparation process for five frequencies separated by 1.5 MHz, with a preparation time of 150 ms.  The first anti-polarisation step used a 300~kHz chirped burn but subsequent steps used 500~kHz to capture broadened components of each tooth as described above. This strategy optimises the optical depth of the AFC teeth, but does produce tails on said teeth. A clean-up sequence (discussed below) could be used to remove the broadened tails of the teeth and the remaining $\kett[-]{5}$ ions.

 The system was probed with strong 200~ns pulses before and after the comb creation, giving the absorption spectrum in Fig. \ref{fig:5toothAFC} (A). The peaks are well approximated by a Gaussian function with a FWHM of 380~kHz. The peak absorption of 18~dB is close to the maximum peak absorption (20~dB) possible on this optical transition (see Fig. \ref{fig:spin_pumped}), which requires complete spin polarisation and perfect anti-polarisation). The remaining $\sim$ 2~dB can be seen in the nearby $\kett[-]{5}_g$. There is a background of $0.51\pm 0.05$~dB only between the teeth and a slight asymmetry in the tooth shape due to acoustic noise in the laser, mentioned above. This noise arises from a 500 Hz mechanical resonance in the external laser diode cavity, causing occasional frequency jumps of up to hundreds of kilohertz.
    
Fig. \ref{fig:5toothAFC} (B) shows the time-domain transmission of the probe pulse. The echo at 660 ns ($(1.5~\mbox{MHz})^{-1}$) after the input represents the fraction of the pulse stored by the AFC. Comparing the intensity of the input reference to the echo gives an efficiency $ \eta = 22 \pm 1\%$.

This efficiency can be compared to that expected for an infinite comb with Gaussian teeth \cite{Lauritzen2011},
\begin{equation}
    \eta = \frac{d^2}{F^2}\exp(-\frac{d}{F})\exp(-\frac{1}{F^2}\frac{\pi^2}{4\ln 2})\exp(-d_0)
    \label{eqn:efficiency}
\end{equation}
where  $d = 18\pm 4$~dB is the peak optical depth of the comb, $F = 3.94$ is the finesse of the comb, defined as the comb spacing divided by the FWHM tooth width, and $d_0 = 1\pm0.2$ dB is the total background absorption. These values predict an efficiency of 25\%, with the difference likely attributed to overfilling of our finite comb, and the slightly non-Gaussian tooth shape.

    \begin{figure}
        \centering
        \includegraphics[width = \linewidth]{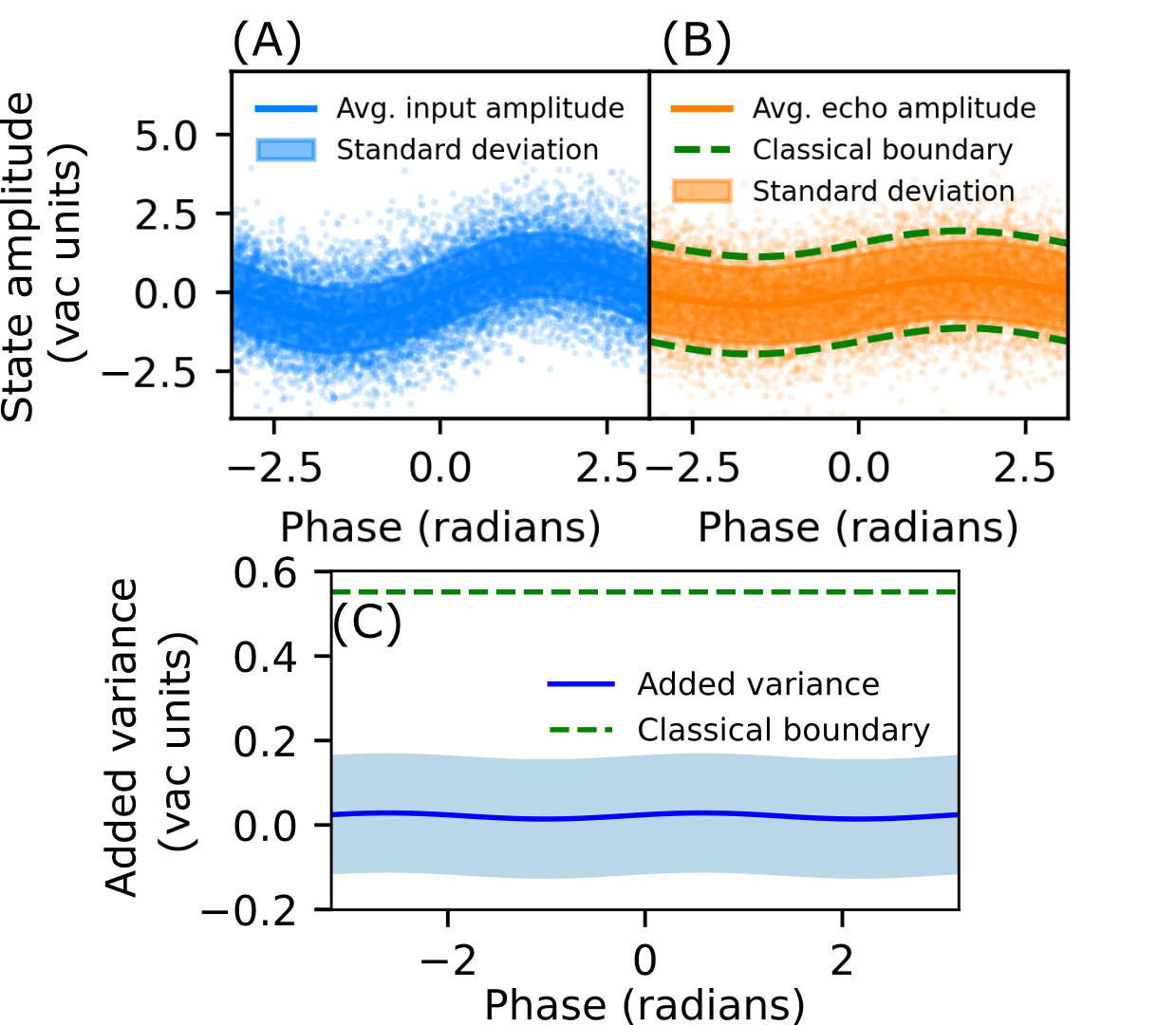}
        \caption{Quadrature values of the input pulse (\textbf{A}) and the echo (\textbf{B}) relative to the phase of the local oscillator, for an average input of 0.8 photons. Shaded band: variance of the pulse amplitude.  (\textbf{C}), variance added to the input pulse by the AFC. Shaded band: 95\% confidence interval.}
        \label{fig:addedVariance}
    \end{figure}

To determine an upper bound for noise added by the storage process, we delayed weak coherent pulses and compared the  noise of the retrieved pulses to vacuum noise.  After each storage event, the transmission of two large, frequency-detuned reference pulses was used to correct for trigger jitter and drifting interferometer phase\cite{Hedges2010}. Each time an AFC was created, $10^4$ storage events were recorded before the entire preparation sequence was repeated. This was repeated 10 times for a total of $10^5$ storage events. By comparing the signal at the detector to shot noise, and accounting for additional losses in the collection path (reflections from cryostat windows, fiber coupling losses and detector quantum efficiency totalling 6.3~dB) we determine the mean photon number for pulses reaching the crystal is $\langle N\rangle = 0.8$.  Figure \ref{fig:addedVariance} (A,B) shows the resulting noise measurements. An upper bound for quadrature noise added due to storage is the difference between the echo's variance and that of an ideal coherent state (i.e. vacuum noise), shown in Fig. \ref{fig:addedVariance} (C). This indicates less than 0.1 unit of vacuum noise is added by the storage with over 95\% confidence. This is within the classical boundary for this efficiency level ($2\eta$) \cite{hedges10}.

These results are an initial demonstration of the quantum memory performance achievable with an optimised spectral preparation strategy in \ErYSO. Because this system can be spin-polarised, we could create a comb with a near-maximal optical depth on a relatively low background. The 22\% efficiency we then observed is two orders of magnitude higher than previous demonstrations of spectral holeburning quantum memories in Er materials with similar storage times \cite{craiciu19a,Saglamyurek2015,askarani19,Lauritzen2011}, with those demonstrations limited by poor spectral holeburning of Er\tplus in low magnetic field, from short life and coherence times.

Despite this large improvement in efficiency, our memory demonstration did not reach the material limits of \ErYSO in the high field regime. Here, we detail what performance can be achieved in this material and the experimental modifications required. Our quantitative predictions are made for an AFC memory but the requirements for other spectral holeburning quantum memories, such as the gradient echo memory, are very similar.

The first step to improve the efficiency is to reduce the 1~dB background absorption. Most (0.9 dB) of this background is due to \Er\ ions and could be removed by a clean-up sequence, the simplest version of which involves burning in between the teeth after comb preparation.  Given the long 188~s lifetime of the $\kett[-]{7}_g$ level relative to the 10 ms excited state life time, we would expect to be able to remove more than 99.9\% of the $^{167}$Er ions, leaving only the I=0 erbium ions (0.08 dB). During this clean up sequence we would also burn away the remaining $\kett[-]{5}_g$ ions (Fig. \ref{fig:5toothAFC}(A)). It is possible to burn away ions in the $\kett[-]{5}_g$ level even if they overlap with the $\kett[-]{7}_g$ in the $\Delta m_I = 0$ band by burning on another $\Delta m_I$ transition, such as the +1, since the ions would not overlap in that band. Applying such a clean-up sequence would allow, in principal, increase the efficiency to 30\%. 

Substantial further improvements in efficiency are then possible using an impedance-matched optical cavity \cite{Afzelius2010, Moiseev10, Jobez2014}. Then, the memory efficiency is limited only by the ratio $\frac{d}{d_0}\approx 225$ of the peak to background absorption. We calculate, using Eq. 5 from \cite{Jobez2014}, that for an AFC finesse of 9, impedance matching with a 100~MHz bandwidth can be achieved using a 27 cm cavity with a finesse of 11. For this setup the predicted efficiency rises to 89\%.

This efficiency is limited by the background absorption of the 8\% $I=0$ isotope impurities. 96.3\% purity \Er\ is routinely available, giving a predicted efficiency of 93\%. Other $\Lambda$ systems with higher $\frac{d}{d_0}$ could also be used to increase the efficiency, such as ``option 2" in Fig. \ref{fig:spin_pumped} (B). 

We now consider the bandwidth. We prepared a 6~MHz wide AFC, and increasing the bandwidth would simply require creating more teeth. The only cost to doing so (provided a more stable laser than we had available) is a linear increase in preparation time, but even a 100 MHz comb could be prepared in 2.5~s, much shorter than the 188 s population lifetime. A 100 MHz comb would have a slightly lower average peak absorption of 16.5~dB and a slightly higher average $I=0$ background of 0.095~dB than the comb prepared here, but $>90$\% efficiency is still possible assuming 96.30\% isotopic purity and an efficient clean-up sequence.

The storage time of the memory can be extended with the use of spin state storage to the order of the coherence time of the spin levels, 1.3~s\cite{Rancic2016}. In this case, the clean-up sequence would need to be extended to clean both the $\kett[-]{7})g$ (memory level) and the $\kett[-]{5}_g$ (spin-state storage level).   Even when considering a memory with a 100 MHz bandwidth, the clean-up sequence is still viable (although at these extreme bandwidths the $\kett[-]{3}_g$ would also need to be cleaned). This is because the preparation processes are limited by the relatively long excited state life time (10ms), rather than laser power. For instance, with our Rabi frequency of 500 kHz, an entire 100 MHz line can be excited in 400 us. Achieving spin-state storage then requires optical $\pi$ pulses applied during optical storage, 660~ns here. Our 500 kHz Rabi frequency was slightly too low for spin shelving, but this could be readily increased by amplifying the 200~$\mu$W laser or reducing the 40~$\mu$m spot size on the crystal\cite{Vivoli2013}. For example, a spot size of  10~$\mu$m is easily achievable with different focusing optics, and an erbium doped fiber amplifier could be used to amplify the laser to 50 mW, increasing the Rabi frequency to $\order{100}$~MHz. Such additions would allow spin state storage pulses with bandwidths comparable to the entire inhomogeneous line.

In conclusion, we have described a spectral preparation strategy to optimise quantum memory performance in systems with resolved, long-lived hyperfine structure and transitions clustered according to \dm. We demonstrate this strategy in \ErYSO at high field and use it to demonstrate a 6~MHz wide AFC with 22\% storage efficiency operating at the quantum level. This is already the most efficient storage demonstrated with atoms in the telecommunications C-band. With further experimental improvements, the same material and techniques are appropriate for high performance quantum memory meeting the requirements of global quantum repeaters\cite{Simon2010, tittel2010}.

\begin{acknowledgments}
This work was supported by the Australian Research Council Centre of Excellence for Quantum Computation and Communication Technology (Grant No. CE170100012) and the Commonwealth of Australia Defence Science and Technology Group.

\end{acknowledgments}

\bibliography{AFC_paper_PRR}

\end{document}